\documentclass{easychair}

\usepackage{doc}
\usepackage{makeidx}
\usepackage{listings}
\usepackage{algorithm}
\usepackage{tikz}
\usepackage{epstopdf}
\usepackage{cite}
\usepackage{subfigure}
\usepackage{enumitem}
\setlist{nolistsep}
\usepackage{titlesec}
\titlespacing\section{0pt}{12pt plus 4pt minus 2pt}{0pt plus 2pt minus 2pt}
\titlespacing\subsection{0pt}{12pt plus 4pt minus 2pt}{0pt plus 2pt minus 2pt}
\titlespacing\subsubsection{0pt}{12pt plus 4pt minus 2pt}{0pt plus 2pt minus 2pt}

\usetikzlibrary{positioning}

\begin{document}

%
\title{Performance Evaluation of Unified Parallel C for\\  Molecular Dynamics}


\titlerunning{Performance Evaluation of  {\Unified Parallel C} for\\  Molecular Dynamics}

%
\author{
Kamran Idrees\inst{1}
\and
    Christoph Niethammer\inst{1}
\and
   Aniello Esposito\inst{2} \\
\and
   Colin W. Glass\inst{1}
}

\institute{
  High Performance Computing Center Stuttgart (HLRS),
  Stuttgart, Germany\\
  \email{idrees@hlrs.de, niethammer@hlrs.de, glass@hlrs.de}
\and
   Cray Inc.,
   Seattle, WA, U.S.A.\\
   \email{esposito@cray.com}\\
 }

\authorrunning{Idrees, Niethammer, Esposito and Glass}

\clearpage

\maketitle

\begin{abstract}
Partitioned Global Address Space (PGAS) integrates the concepts of shared memory programming and the control 
of data distribution and locality provided by message passing into a single parallel programming model. 
The purpose of allying distributed data with shared memory is to cultivate a locality-aware shared memory paradigm. 
PGAS is comprised of a single shared address space, which is partitioned among threads. Each thread has a portion 
of the shared address space in local memory and therefore it can exploit data locality by mainly doing 
computation on local data. 
\newline Unified Parallel C (UPC) is a parallel extension of ISO C and an implementation of the PGAS model. 
In this paper, we evaluate the performance of UPC based on a 
real-world scenario from Molecular Dynamics.
\end{abstract}

\setcounter{tocdepth}{2}

%
%

\pagestyle{empty}

\section{Introduction}
\label{sect:introduction}

\label{sec:1}
Partitioned Global Address Space (PGAS) is a locality-aware distributed shared memory model for Single Program 
Multiple Data (SPMD) streams. PGAS unites the concept of shared memory programming and distributed data. 
It provides an abstraction of Global Shared Address Space, where each thread can access any memory location 
using a shared memory paradigm. The Global Shared Address Space is formed by 
integrating the portions of the memories on different nodes and the low level communication involved for 
accessing remote data is hidden from the user. Unified Parallel C (UPC) is an implementation of the PGAS model. 
The low-level communication in UPC is implemented using light-weight Global-Address Space Networking (GASNet).
UPC benefits from the brisk one-sided communication provided by GASNet and thus has a performance advantage 
over message passing ~\cite{Yelick:2007:PPU:1278177.1278183}.

Molecular Dynamics simulates the interactions between molecules ~\cite{allen2004introduction}. 
After the system is initialized, the
forces acting on all molecules in the system are calculated.
Newton's equations of motion are integrated to advance the
positions and velocities of the molecules. The simulation is
advanced until the computation of the time evolution of the
system is completed for a specified length of time. 

In this paper, we evaluate the intra- and inter-node performance of
UPC based on a real-world application from Molecular Dynamics, compare it intra-node with OpenMP
and show the necessity for manual optimizations by the programmer in order to achieve good performance.

\section{Unified Parallel C}
\label{sec:2}
Unified Parallel C (UPC) is a parallel extension of ISO C. It is a distributed shared memory programming model that runs in a 
SPMD fashion, where all threads execute the main program or function. Using UPC constructs, each thread 
can follow a distinct execution path to work on different data. UPC threads run independently of each other, the
only implied synchronization is at the beginning and at the end of the main function ~\cite{el2005upc}. It is the
responsibility of the programmer to introduce necessary synchronization when  shared data is accessed by more than one 
thread. Apart from a global shared address space, UPC also provides private 
address space for each thread. The private address space is only accessible by the thread whichs owns it. 
This allows a programmer to intelligently allocate the data in private and shared address spaces. 
Data which remains local to a thread should be allocated on the private address space. Whereas data which needs to be 
accessed by multiple UPC threads, should be allocated on the portion of the shared address space of the thread 
doing most computation on it ~\cite{el2005upc}.

UPC accommodates several constructs which allow to allocate  data and the thread with affinity to it on the same physical node. 
UPC also provides constructs to check the locality of  data. The programmer needs to identify data as local in order to access
it with a local pointer.

UPC utilizes a source to source compiler. The
source to source compiler translates UPC code to ANSI C code (with additional code for
communication to access remote memory, which is hidden from the user) and links to the UPC run-time system.
The UPC run-time
system can examine the shared data accesses and perform communication optimizations 
~\cite{Yelick:2007:PPU:1278177.1278183}~\cite{el2005upc}.

\section{Molecular Dynamics Code}
\label{sec:3}
We ported our in-house Molecular Dynamics code CMD,
developed for basic research into high performance
computing. CMD features multiple MD data structures,
algorithms and parallelization strategies and thus allows for
quantitative comparisons between them. Two widely
used data structures are implemented - with corresponding
algorithms - for the computation of interactions between
molecules in the system, ``BasicN2'' and 
``MoleculeBlocks''. The MoleculeBlocks code has been ported to 
UPC.

MoleculeBlocks implements a linked cell approach, where the domain is spatially
decomposed into cells (of the size cut-off radius) and then the
molecules are distributed among these cells. In this algorithm,
the distances between the molecules are computed only intra-cell and for neighboring cells. Furthermore,
Newton's 3rd law of motion is used to reduce the compute
effort by half. Figure ~\ref{example} shows an example of the
MoleculeBlocks algorithm for a 2D domain space. When the interaction between a pair of
molecules is computed, the resulting force is written to both
molecules. Thus, the centered cell (dark gray), as shown in figure ~\ref{example}, modifies the
forces of its own molecules and molecules of its right and lower
neighbor cells (gray).
Although the use of Newton’s 3rd law lessens
the computational effort, it raises the requirements regarding
synchronization in order to avoid race conditions.

\begin{figure}
\centering
\parbox{2.5in}{%
\begin{tikzpicture}[scale=0.68]
\centering
\path[draw, thick, fill=gray] (0,-1) rectangle (1,2);
\path[draw, thick, fill=gray] (-1,-1) rectangle (0,0);
\path[draw, thick, fill=darkgray] (-1,0) rectangle (0,1);

\path[draw, thick, fill=lightgray] (-2,-1) rectangle (-1,2);
\path[draw, thick, fill=lightgray] (-1,1) rectangle (0,2);

\foreach \i in {-3,-1,1}{
\foreach \j in {-3,-1,1}{

\draw [thick, fill] (\i +1.7 +0.1*\j,\j+ 1.4) circle [radius=.12]; 
\draw [thick, fill] (\i +0.67,\j+ 1.75+0.1*\i) circle [radius=.12]; 
\draw [thick, fill] (\i +1.3+0.05*\j,\j+ .3) circle [radius=.12]; 
\draw [thick, fill] (\i +0.3-0.01*\j,\j+ 0.8+0.01*\j) circle [radius=.12]; 
}
}

\draw [thick] (-1+0.67, .75-0.1) circle [radius=1]; 

\draw[step=1.0, thick] (-3,-3) grid (3,3);

\end{tikzpicture}
\caption{Calculation of interaction between molecules using the \texttt{MoleculeBlocks} algorithm.}
\label{example}
}%
\qquad
\begin{minipage}{2.5in}
\begin{tikzpicture}[every node/.style={minimum size=0.5cm},on grid,scale=0.50]
\begin{scope}[every node/.append style={yslant=-0.5},yslant=-0.5]
  \shade[right color=gray!10, left color=black!50] (0,0) rectangle +(4,4);
  \draw (0,0) grid (4,4);
\end{scope}
\begin{scope}[every node/.append style={yslant=0.5},yslant=0.5]
  \shade[right color=gray!70,left color=gray!10] (4,-4) rectangle +(4,4);
  \draw (4,-4) grid (8,0);
\end{scope}
\begin{scope}[every node/.append style={
    yslant=0.5,xslant=-1},yslant=0.5,xslant=-1
  ]
  \shade[bottom color=gray!10, top color=black!80] (8,4) rectangle +(-4,-4);
  \draw (4,0) grid (8,4);
\end{scope}
\end{tikzpicture}
\caption{Domain is spatially decomposed into cells and distributed among threads in a spatially coherent manner.}
\label{fig:1}
\end{minipage}
\end{figure}

\section{Porting Molecular Dynamics Code to UPC}
\label{sec:4}
For MoleculeBlocks, the system of molecules (called Phasespace) is spatially decomposed into cells 
where each cell contains a number of molecules (as shown in figure ~\ref{fig:1}). The cells are then distributed among 
the UPC threads in a spatially coherent manner (as opposed to the default round-robin fashion) to reduce the 
communication overhead between the UPC threads. The CMD simulation code is comprised of two parts, (i) phasespace initialization 
\& grid generator and (ii) main simulation loop.

\subsection{Phasespace Initialization \& Grid Generator}
Phasespace initialization involves allocation of the memory dynamically on the global shared space for molecules
and cells.
A routine for performing the transformation (or mapping) of a spatially coherent cell indices (i,j,k) 
to consecutive integers (cell IDs) is introduced in the code, which allows UPC shared array to 
distibute spatially coherent cells among UPC threads by blocking consecutive cell IDs.
The Grid generation routine initializes the positions and velocities of the molecules and adds them to
the phasespace.

\subsection{Main Simulation Loop}
All the routines inside the main simulation loop have been implemented in UPC as locality-aware algorithms. 
Each thread only calculates the 
different parameters for the cells which reside in its portion of shared space. Due to manual optimization, each
thread accesses its cells using local pointer instead of pointer-to-shared.

The Lennard-Jones Force Calculation routine is the most compute intensive routine of the simulation. Here, 
each thread computes the inter-molecular interactions, for the cells residing in its portion of 
shared space and with the neighbor cells which may reside locally or remotely. 
The synchronization among threads is provided through a locking strategy. 
We have tested different locking granularities, which are ``lock per molecule'' and ``lock per cell''. Furthermore, 
we have also implemented pre-fetching and copy-at-once strategies to reduce the communication between UPC threads. 
This has a major effect when the UPC code is scaled beyond a single node. With pre-fetching and copy-at-once 
strategies, a UPC thread pre-fetches the values (of positions and forces) of the 
molecules of its neighbor cell if the neighbor cell is remote (i.e. does not reside in its local portion of the shared space) 
to its private space. The function of pre-fetching is implemented using the upc\_memget routine.
Once a UPC thread has pre-fetched the data, it computes all interactions between the molecules 
of the two cells and then copies all the calculated forces to the neighbor cell in one go using the upc\_memput routine. 

In order to calculate the global value of 
parameters (e.g potential energy), the coordination among threads is done using the reduction
function upc\_all\_reduceT available in the collective library of UPC.

\section{Hardware Platforms}
\label{sec:5}

The presented benchmarks have been produced on a Cray XE6 system
and a Cray XC30 system. The code was
built by means of the Berkeley UPC compiler (version 2.16.2) on both systems. 
The cray compiler had some performance issues which are under investigation.
The nodes of the XE6 system feature two AMD Interlagos processors (AMD Opteron(TM) Processor 6272 clocked at 2.10GHz) 
with 16 integer cores each. Two integer cores share a floating point unit. 
On the other hand, the compute nodes of the XC30 system contain two Intel
Sandybridge processors (Intel(R) Xeon(R) CPU E5-2670 0 clocked at 2.60GHz) with
8 cores and 8 hyperthreads each. For the present benchmarks no hyperthreads have been used. 


For a full documentation and technical
specifications of the hardware platforms, the reader is referred to the online
material\footnote{www.cray.com/Products/Computing/XE/Resources.aspx,
   www.cray.com/Products/Computing/XC/Resources.aspx}.

\section{Evaluation}
\label{sec:6}
The UPC implementation of CMD is evaluated for the cut-off radius of 3  with the following strategies (a lock 
always locks the entire cell).

\begin{enumerate}
  \item {\emph{Lock per molecule (LPM)} - Acquire lock for each molecule-molecule interaction}
  \item {\emph{Lock per cell (LPC)} - Acquire lock for each cell-cell interaction}
  \item {\emph{Lock per cell plus prefetching (LPC+)} - Same as lock per cell but with pre-fetching and copy-at-once strategies}
\end{enumerate}

The cut-off radius determines the maximum distance for evaluating molecule-molecule interactions. Increasing the cut-off
will result in more interactions per molecule and therefore more computational effort. The cell size is equal to 
the cut-off radius.

In the first UPC implementation of CMD, we did not perform the manual pointer optimizations. The shared address space 
was always accessed using pointer-to-shared irrespective of the fact whether the data accessed by a UPC thread is local 
or remote. The test cases of 500, 3,000 and 27,000 molecules without pointer optimizations executed 10 times 
slower than the current version which incorporates manual pointer optimizations.

The rest of the evaluation is based on the version with manual optimization, a cut-off radius of 3 and with 6,000,000 molecules. Thus, all scaling benchmarks shown here are based on strong scaling.
The evaluation metrics are explained in the following 
subsection. 

\subsection{Evaluation Metrics}
The UPC implementation of CMD is evaluated for all three strategies described above, on the basis of intra- and inter- node 
performance. For each case, we have taken the average execution time for five program runs.

\vspace{0.5em}
{\bf Intra-Node Performance} 
The intra-node performance compares  the UPC and OpenMP implementations
of CMD on a single node.

\vspace{0.5em}
{\bf Inter-Node Performance} 
The inter-node performance shows the scaling 
on multiple nodes. 
Under populated nodes are used for inter-node results when the number of UPC threads are less than the total 
available CPU count of 4 nodes. The threads are always equally distributed among the nodes.

\subsection{Results}
Here we show the execution time of CMD both intra- and inter-node, and compare the execution time
with varying locking granularities.

\begin{figure}
\centering
\parbox{2.5in}{%
\includegraphics[width=2.5in]{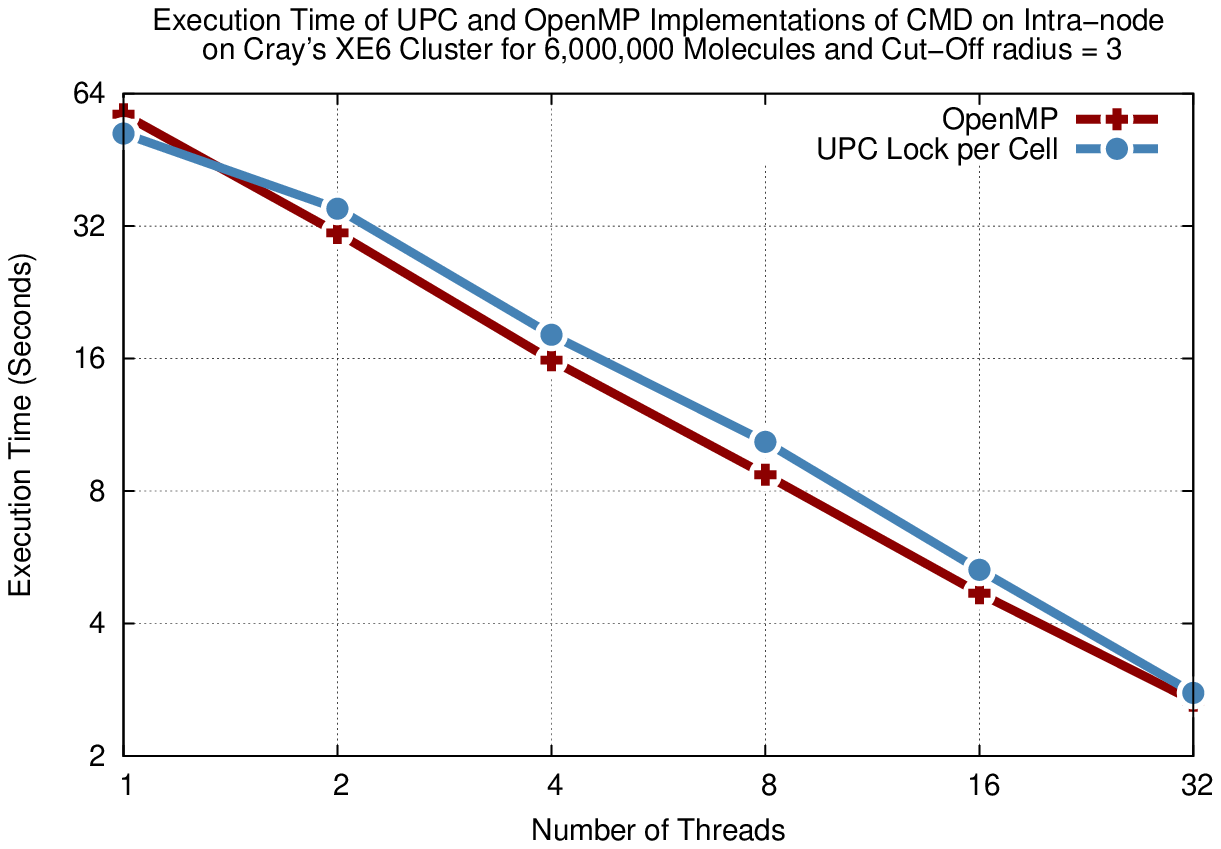}
\caption{Comparison of intra-node execution time of UPC and OpenMP implementations of CMD on a Cray XE6.}
\label{fig:ompupc}      
}%
\qquad
\begin{minipage}{2.5in}
\includegraphics[width=2.5in]{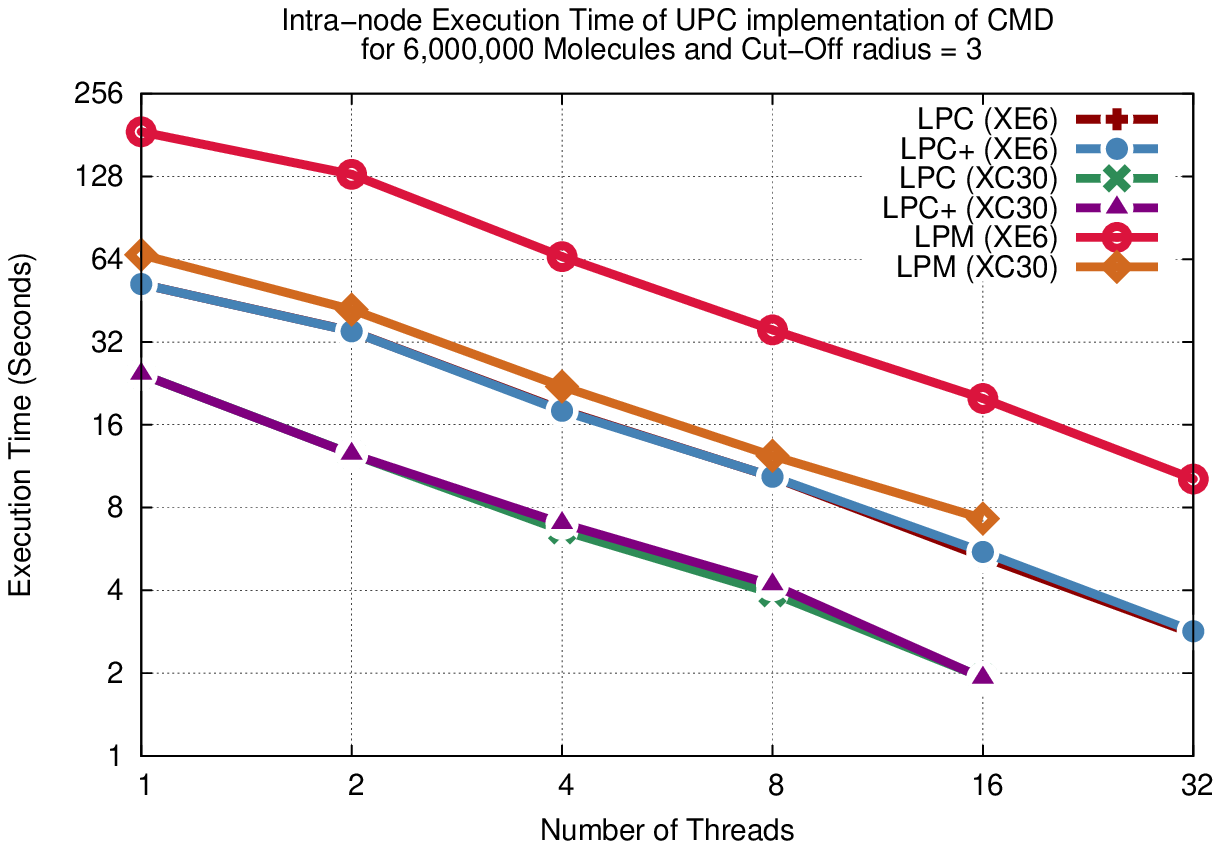}
\caption{Intra-node execution time of UPC implementation of CMD with different synchronization strategies.}
\label{fig:intra}
\end{minipage}

\end{figure}

\vspace{0.5em}
{\bf Intra-Node Performance} 
Figure ~\ref{fig:intra} shows the intra-node benchmark for the execution time of UPC implementation of CMD. 
Clearly, the lock per cell strategy is superior to the lock per molecule. Pre-fetching and copy-at-once has 
no significant impact on intra-node performance. Figure ~\ref{fig:ompupc} compares intra-node performance achieved 
with OpenMP and UPC. UPC performs similarly to OpenMP on a single node. 
This is a satisfactory result, as the aim is not to provide a better shared memory parallelization, 
but to use a shared memory paradigm for distributed memory. 
Having a comparable performance as OpenMP is a good basis.

\begin{figure}
\centering
\parbox{2.5in}{%
\includegraphics[width=2.5in]{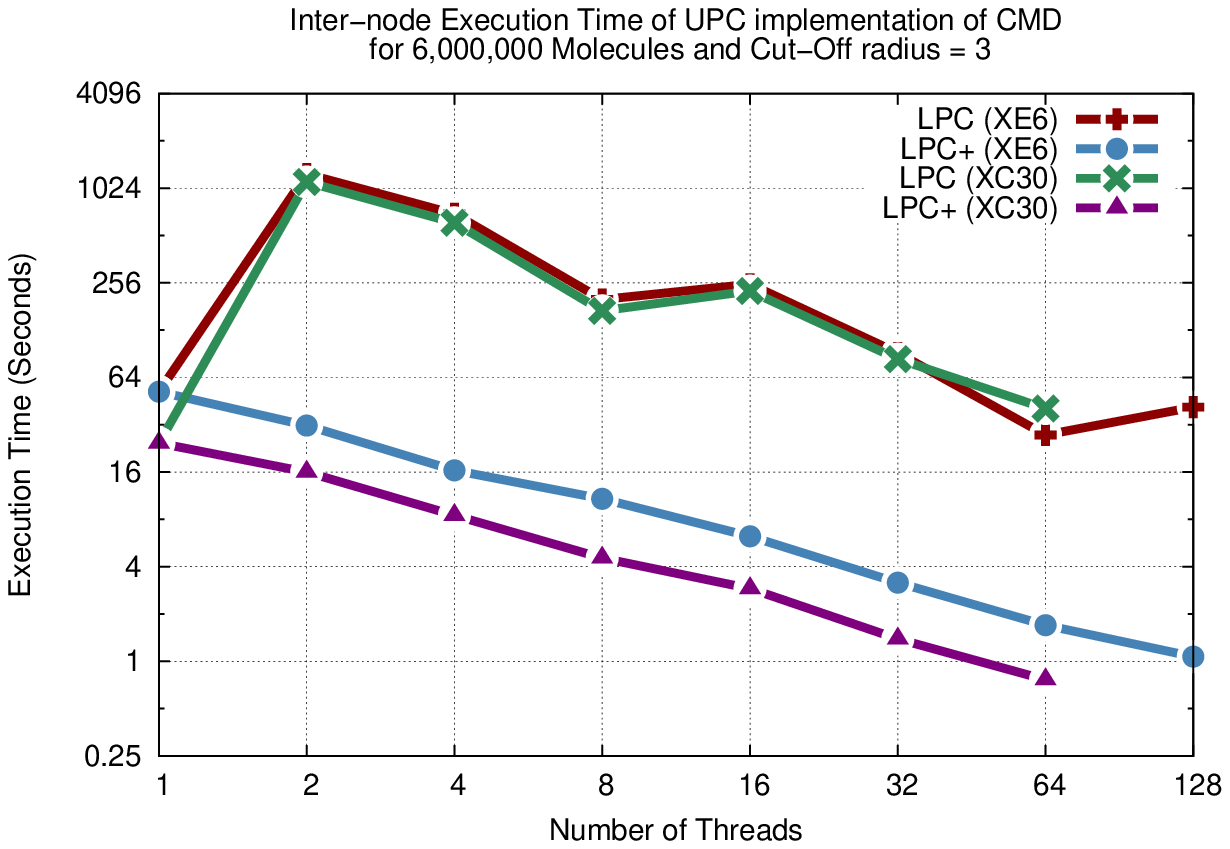}
\caption{Inter-Node Execution Time of the UPC implementation with different locking strategies and cut-off radius of 3.}
\label{fig:inter}      
}%
\qquad
\begin{minipage}{2.5in}
\includegraphics[width=2.5in]{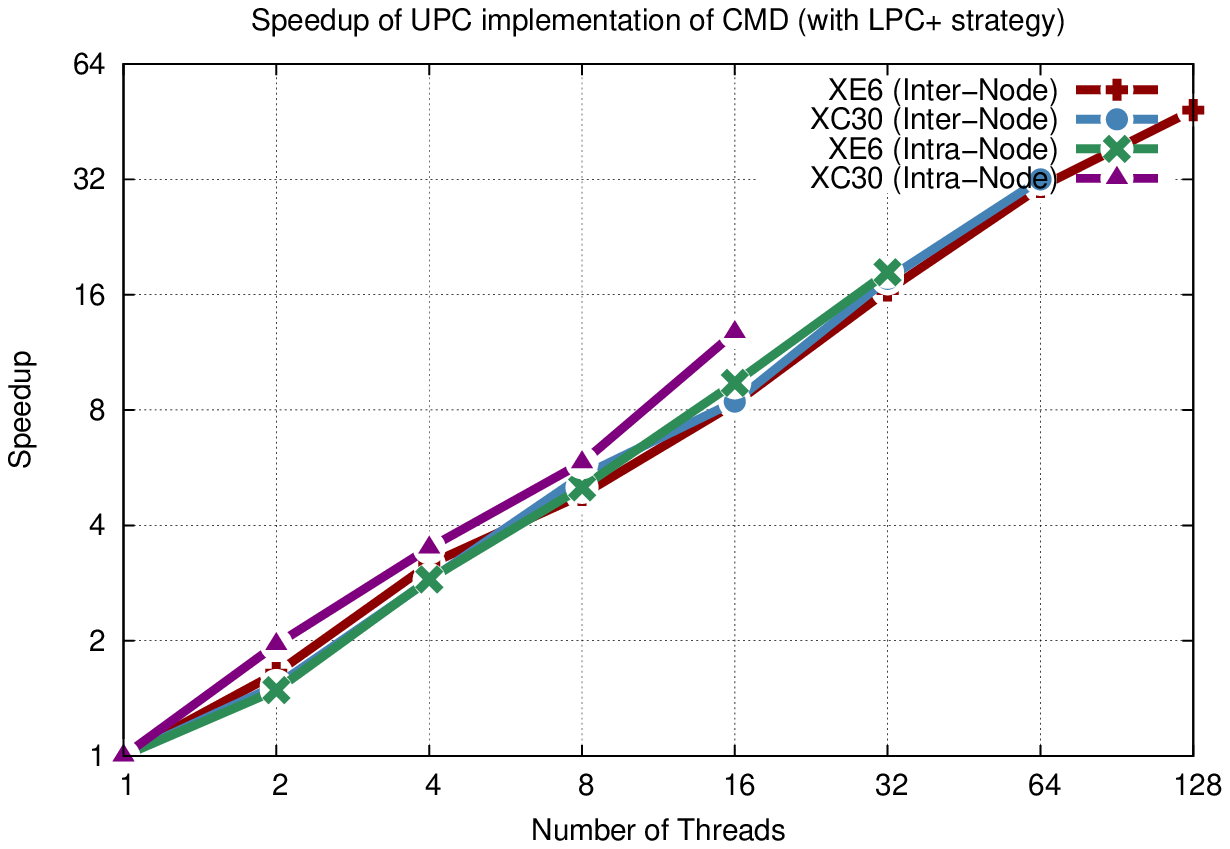}
\caption{Speedup of UPC implementation of CMD with Lock per cell plus pre-fetching strategy on intra- and inter-node.}
\label{fig:speedup}    

\end{minipage}
  
\end{figure}

\vspace{0.5em}
{\bf Inter-Node Performance} 
Figure ~\ref{fig:inter} shows the inter-node performance achieved with UPC, using 
LPC and LPC+. The LPM strategy is disregarded due to its inferior intra-node performance. 
As can be seen, the LPC strategy shows very poor inter-node performance (2 or more threads). 
The execution time jumps by a factor of 20+, as soon as inter-node communication comes into play. 
However, the LPC+ strategy shows a solid scaling behaviour, making this implementation 
competitive even for production runs. Figure ~\ref{fig:speedup} shows the speedup of CMD with LPC+ strategy on 
intra- and inter-node, for both XE6 and XC30 clusters.

\section{Conclusion}
\label{sec:7}
As we have shown in this paper, it is possible to implement a competitive distributed memory parallelization using UPC. 
However, the implementation is far from trivial. There are many pitfalls, leading to significant performance 
degradations and they are not always obvious.

The first and most elusive problem is the use of pointer-to-shared for local data accesses. As a programmer, 
one would expect UPC compilers or the run-time to automatically detect local data accesses and 
perform the necessary optimization. However, this is not the case and the performance degradation for our 
use case was a staggering factor 10. This suggests that with the currently available compilers,
manual pointer optimization (using local C pointers when a thread has affinity to 
the data) is mandatory.

The second issue is not discussed in detail here. In a nutshell: the default round robin distribution of 
shared array elements leads to significant communication traffic in this scenario. 
Manual optimization was necessary, essentially replacing round robin with a spatially coherent distribution. 
Thus, the programmer needs to keep the underlying distributed memory architecture in mind.

The third and maybe most disturbing problem is related to communication granularity. 
The LPM and LPC strategies represent the way one traditionally would approach a shared memory parallelization. 
As the data is available in shared memory, there is no need to pre-fetch it or to package communication. 
However, these approaches fail completely when utilized for distributed memory, as can be seen in figure ~\ref{fig:inter}.

The good news is, all the above problems can be solved, the bad news is: it requires the programmer to think 
in terms of distributed memory parallelization. However, this is not the driving idea behind PGAS.

In our view, in order for PGAS approaches to prosper in the future, these issues have to be addressed.
Only if better data locality, data distribution and communication pooling is provided automatically by 
the compiler or run-time will programmers start seeing true benefit. 
The required information for such improved automatic behaviour is available to the compiler, 
as the parallelization is achieved purely through UPC data structures and routines.

\paragraph{Acknowledgments}
This work was supported by the project \emph{HA} which is funded by the German Research Foundation (DFG) 
under the priority programme "Software for Exascale Computing - SPPEXA" (2013-2015) and the EU project \emph{APOS} 
which was funded as part of the European Commission’s Framework 7.
%
\label{sect:bib}
\bibliographystyle{plain}
\bibliography{easychair}

\end{document}